\documentclass[11pt]{article}
\usepackage{latexsym}  
\usepackage{color}       
\usepackage{graphicx}    
\usepackage{graphicx,amsmath,amssymb}
\usepackage{amssymb}
\usepackage{amsmath}

\begin{document}

\hspace*{9cm}{OU-HET-711/2011, MISC-2011-11}

\begin{center}
{\Large\bf 
  Neutrino Mass Matrix with No Adjustable Parameters  }

\vspace{4mm}

\vspace{4mm}
{\bf Yoshio Koide$^a$ and Hiroyuki Nishiura$^b$}

${}^a$ {\it Department of Physics, Osaka University, 
Toyonaka, Osaka 560-0043, Japan} \\
{\it E-mail address: koide@het.phys.sci.osaka-u.ac.jp}

${}^b$ {\it Faculty of Information Science and Technology, 
Osaka Institute of Technology, 
Hirakata, Osaka 573-0196, Japan}\\
{\it E-mail address: nishiura@is.oit.ac.jp}

\date{\today}
\end{center}

\vspace{3mm}
\begin{abstract}
On the basis of the so-called ``yukawaon" model, 
we found out a special form of the neutrino mass matrix $M_\nu$ 
which gives reasonable predictions.  
The $M_\nu$ is given by a multiplication form made of 
charged lepton mass matrix $M_e$ and up-quark mass 
matrix $M_u$. This $M_\nu$ has no adjustable parameters   
except for those in $M_e$ and $M_u$. 
Here, $M_e$ and $M_u$ are described by one parameter
$a_e$ (real) and two parameters $a_u$ (complex), 
respectively, and those parameters are constrained 
by their observed mass ratios. With this form of $M_\nu$,  
in spite of having only three parameters, 
the $M_\nu$ can give reasonable predictions  
$\sin^2 2\theta_{atm} \simeq 0.99$, $\sin^2 2\theta_{13} \simeq 0.015$, 
$\Delta m^2_{21}/\Delta m^2_{32} \simeq 0.030$, 
$\langle m_{ee}\rangle \simeq 0.0039$ eV, and so on, 
by using observed values of $m_e/m_\mu$, $m_\mu/m_\tau$, $m_c/m_t$, 
and $\sin^2 \theta_{solar}$ as input values.  
\end{abstract}

PACS numbers:  14.60.Pq,  
  11.30.Hv, 
  12.60.-i, 

\vspace{3mm}



\section{Introduction}

The observed masses and mixings of the quarks and leptons
will provide a promising clue to a unified understanding
of those fundamental particles. 
For such the purpose, not only investigating 
a theoretical model, but also searching for phenomenological 
mass matrix seem to be still effective.
As one of phenomenological mass matrix models of the quarks and leptons, 
the so-called ``yukawaon" model \cite{yukawaon}
(a kind of ``flavon" models \cite{flavon}) has been proposed. 
Here, the ``effective" Yukawa coupling constants $Y_f^{eff}$ are given
by 
$$ 
(Y_f^{eff})_{ij} = \frac{y_f}{\Lambda} \langle (Y_f)_{ij} \rangle 
\ \ \ \ \ (i,j=1,2,3),
\eqno(1.1)
$$ 
$\Lambda$ is an energy scale of 
the effective theory, and   
$\langle Y_f \rangle$ are vacuum expectation value (VEV) matrices  
of scalar fields $Y_f$ with $3\times 3$ components. 
(Hereafter we call fields $Y_f$ ``yukawaons".) 
The most characteristic point in the yukawaon model is that 
all VEVs $\langle Y_f \rangle$
are described in terms of 
only one fundamental VEV matrix  
$\langle \Phi_e \rangle \propto {\rm diag}(\sqrt{m_e}, \sqrt{m_\mu},
\sqrt{m_\tau})$.
For example, in a yukawaon model with an O(3) family symmetry 
\cite{O3_PLB09,N-K_PRD11}, 
the charged lepton, neutrino, up-quark, and down-quark
mass matrices $M_e$, $M_\nu$, $M_u$ and $M_d$ are given by

$$
M_e \propto \langle Y_e \rangle \propto \langle \Phi_e \rangle
\langle \Phi_e \rangle ,
\eqno(1.2)
$$
$$
M_\nu = m_D M_R^{-1} m_D^T ,
\eqno(1.3)
$$
$$
m_D \propto  \langle Y_e \rangle ,  \ \ \ \ 
M_R \propto \langle Y_R \rangle +m_{0\nu}^{-1}
\langle Y_e \rangle \langle Y_e \rangle ,
\eqno(1.4)
$$
$$
\langle Y_R \rangle \propto 
\langle \Phi_u \rangle \langle P_u \rangle \langle Y_e \rangle 
+\langle Y_e \rangle \langle P_u \rangle \langle \Phi_u \rangle
+\xi_\nu \left( 
\langle \Phi_u \rangle \langle P_u \rangle \langle Y_e \rangle 
+\langle Y_e \rangle \langle P_u \rangle \langle \Phi_u \rangle
\right) .
\eqno(1.5)
$$
$$
M_u \propto \langle Y_u \rangle \propto \langle \Phi_u \rangle
\langle \Phi_u \rangle ,
\eqno(1.6)
$$
$$
\langle \Phi_u \rangle \propto \langle \Phi_e \rangle
(\langle E\rangle + a_u \langle X \rangle ) \langle \Phi_e \rangle ,
\eqno(1.7)
$$
$$
M_d \propto \langle Y_d \rangle \propto \langle \Phi_e \rangle
(\langle E \rangle + a_d \langle X \rangle) \langle \Phi_e \rangle ,
\eqno(1.8)
$$
where
$$
\langle E \rangle = v_E {\bf 1} = v_E \left(
\begin{array}{ccc}
1 & 0 & 0 \\
0 & 1 & 0 \\
0 & 0 & 1
\end{array} \right) ,
\ \ \ 
\langle X \rangle = v_X S_3 \equiv \frac{1}{3} v_X \left(
\begin{array}{ccc}
1 & 1 & 1 \\
1 & 1 & 1 \\
1 & 1 & 1
\end{array} \right) ,
\eqno(1.9)
$$
in the diagonal basis of $M_e$, while 
$\langle P_u \rangle$ is given by a form
$$
\langle P_u \rangle_u \propto \left(
\begin{array}{ccc}
1 & 0 & 0 \\
0 & -1 & 0 \\
0 & 0 & 1
\end{array} \right) ,
\eqno(1.10)
$$
in the diagonal basis of $M_u$.
 
In this paper, we will find  out a special form of the neutrino 
mass matrix which is 
compatible with the observed neutrino data in spite of 
having no adjustable parameters. 
The form will be obtained along the lines of the yukawaon model 
by changing the structure of $\langle Y_e \rangle$ from Eq.(1.2).  

The neutrino mass matrix is still given by the form of seesaw type, 
$M_\nu = m_D M_R^{-1} m_D^T$ with $m_D \propto  \langle Y_e \rangle$,
but the Majorana neutrino mass matrix $M_R$ is changed into 
a simple form
$$
M_R \propto \langle Y_R \rangle \propto 
\langle \Phi_u \rangle \langle Y_e \rangle 
+\langle Y_e \rangle \langle \Phi_u \rangle .
\eqno(1.11)
$$
Here, we assume that the charged lepton mass matrix $M_e$ is
given by 
$$
\langle {Y}_e \rangle \propto \langle \bar{\Phi}_0 \rangle
(\langle E' \rangle + a_e \langle X_2 \rangle) 
\langle \bar{\Phi}_0 \rangle ,
\eqno(1.12)
$$
differently from Eq.(1.2), 
and we also redefine $\langle \Phi_u \rangle$ as 
$$
\langle \Phi_u \rangle \propto \langle \Phi_0 \rangle
(\langle E\rangle + a_u \langle X_3 \rangle ) \langle \Phi_0 \rangle ,
\eqno(1.13)
$$
where $\langle E' \rangle = \langle E \rangle = v_E {\bf 1}$ and
$$
\langle X_2 \rangle = v_X S_2 \equiv \frac{1}{2} v_X \left(
\begin{array}{ccc}
1 & 1 & 0 \\
1 & 1 & 0 \\
0 & 0 & 0
\end{array} \right) ,
\ \ \ 
\langle X_3 \rangle = v_X S_3 \equiv \frac{1}{3} v_X \left(
\begin{array}{ccc}
1 & 1 & 1 \\
1 & 1 & 1 \\
1 & 1 & 1
\end{array} \right) ,
\eqno(1.14)
$$

Note that there are no $\xi_\nu$ term, no $m_{0\nu}^{-1}$ term 
and no $\langle P_u \rangle$ in Eq.(1.11), i.e. $M_\nu$ is 
simply given by
$$
M_\nu \simeq k_\nu \left( M_e^{-1} M_u^{1/2} + M_u^{1/2} M_e^{-1} \right).
\eqno(1.15)
$$
In other words, there is no adjustable parameter in  the 
present neutrino mass matrix, except for $a_e$ in $M_e$ and 
$a_u$ in $M_u^{12}$ (i.e. $\langle \Phi_u \rangle$). 
The purpose of the present paper is not to derive the mass 
matrix forms (1.11) - (1.13) theoretically, but to demonstrate 
that the phenomenological neutrino mass matrix (1.15) with
Eqs.(1.11) - (1.13) can be compatible with the present neutrino 
data in spite of quite few parameters.
We predict $\sin^2 2\theta_{atm} \simeq 0.99$, 
$\sin^2 2\theta_{13} \simeq 0.015$, 
$\Delta m^2_{21}/\Delta m^2_{32} \simeq 0.030$, 
$\langle m_{ee}\rangle \simeq 0.0039$ eV, and so on, 
by using observed values of $m_e/m_\mu$, $m_\mu/m_\tau$, 
and $\sin^2 \theta_{solar}$ as input values. 
In this paper, for simplicity, we do not discuss the 
down-quark mass matrix $M_d$ and the Cabibbo-Maskawa- Kobayashi 
\cite{CKM} (CKM) mixing.

In the next section, superpotentials for yukawaons and the 
assignments of the fields in the present U(3) yukawaon model 
are investigated.
In Sec.~3, numerical results of the model are discussed.
Sec.~4 is devoted to the concluding remarks.
In Appendix A, $R$-charge assignments are discussed. 
In Appendix B, we present a rotation matrix which transforms 
$\langle X_3 \rangle$ into $\langle X_2 \rangle$.


\section{Superpotential}

In this section, we give superpotentials for the yukawaons and 
the assignments of the fields in the present U(3) yukawaon model.

In the yukawaon model, the order of the fields is important. 
Therefore, in this paper, let us assume a U(3) family symmetry 
instead of O(3) and  denote fields ${\bf 6}^*$ and ${\bf 6}$ of U(3) as 
$\bar{A}$ and $A$, respectively. 
(Therefore, it should be noted that a term $\bar{A} B \bar{C}$ is allowed, but
$\bar{A}\bar{C} B$ and $B \bar{A}\bar{C}$ are forbidden.) 
In the U(3) model, for example, the relation (1.6) is re-expressed as
$$
(M_u)^{ij} \propto \langle \bar{Y}_u^{ij} \rangle \propto 
\langle \bar{\Phi}_u^{ik} \rangle
\langle {E}^u_{kl} \rangle \langle \bar{\Phi}_u^{lj} \rangle ,
\eqno(2.1)
$$
with $\langle {E}^u \rangle = {v}_{E} {\bf 1}$.
In order to distinguish each yukawaon from other yukawaons,
although we assumed U(1)$_X$ charge in the O(3) model 
\cite{O3_PLB09,N-K_PRD11}, 
in this U(3) model, we assume only $R$ charge conservation 
instead of U(1)$_X$ charge conservation.  
For the right handed neutrino sector ($\bar{Y}_R$), 
it should be noted that we cannot add 
$\bar{Y_e} \bar{Y}_e$ term to $\bar{Y}_R$ as in Eq.(1.4).
In the old model, we assigned the U(1)$_X$ charges $Q_X$
only for gauge singlet fields, e.g. $Q_X(\ell)=Q_X(H_d)$,
i.e. $Q_X(Y_e)=-Q(e^c)$.
Besides, we assumed $Q_X(e^c)=Q(\nu^c)$ in order to 
build a model without $Y_\nu$.
Therefore, we could obtain $Q_X(Y_R)=Q_X(Y_e Y_e)$ in 
the old model.
However, in this U(3) model, we cannot obtain 
$R(Y_R)=R(\bar{Y}_e E \bar{Y}_e)$.
Besides, we cannot introduce a $\xi_\nu$ term such as 
in Eq.(1.5).

We assume the following superpotential 
$W= W_Y + W_e + W_d + W'_u + W_u +W_R + W_E$:
$$
W_Y = \frac{y_e}{\Lambda} {\ell}_i \bar{Y}_e^{ij} e^c_j H_d 
+ \frac{y_\nu}{\Lambda} {\ell}_i \bar{Y}_e^{i j}
\nu^c_j H_u 
+\lambda_R \nu^c_i \bar{Y}_R^{ij} \nu^c_j 
$$
$$
+ \frac{y_u}{\Lambda} u^{c}_i \bar{Y}_u^{ij} q_j H_u 
+ \frac{y_d}{\Lambda} d^{c}_i \bar{Y}_d^{ij} q_j H_d 
+ \mu_H H_u H_d,
\eqno(2.2)
$$
$$
W_e = \mu_e {\rm Tr}[ \bar{Y}_e \Theta^e ] +
\frac{\lambda_e}{\Lambda} {\rm Tr}[ \bar{\Phi}_0 (E'+ 
a_e X_2) \bar{\Phi}_0 \Theta^e ],
\eqno(2.3)
$$
$$
W_d = \mu_d {\rm Tr}[ \bar{Y}_d 
{\Theta}^d] +\frac{\lambda_d}{\Lambda} {\rm Tr}[ 
\bar{\Phi}_0 ( E + a_d e^{i\alpha_d} X_3 ) \bar{\Phi}_0
{\Theta}^d ] ,
\eqno(2.4)
$$
$$
W'_{u} =\mu'_u
{\rm Tr}[\bar{\Phi}_u \Theta^{u\prime}]
+\frac{\lambda^{\prime}_u}{\Lambda} {\rm Tr}[
\bar{\Phi}_0 (E +a_u e^{i\alpha_u} X_3) \bar{\Phi}_0 
\Theta^{u \prime} ]  ,
\eqno(2.5)
$$
$$
W_u = \mu_u {\rm Tr}[\bar{Y}_u {\Theta}^u ] +
\frac{\lambda_{u}}{\Lambda} {\rm Tr}[\bar{\Phi}_u {E}^u
\bar{\Phi}_{u} {\Theta}_u] 
+   \frac{\lambda_{0u}}{\Lambda^{5}} {\rm Tr} [
  (\bar{E}_u E^u)^{3} \bar{E}_u \Theta^u ]  ,
\eqno(2.6)
$$
$$
W_R = \mu_R {\rm Tr}[ \bar{Y}_R  {\Theta}^R ] +
\frac{\lambda_R}{\Lambda}{\rm Tr}[ \left( \bar{\Phi}_u  
E^u \bar{Y}_e  + \bar{Y}_e {E}^u \bar{\Phi}_u \right) {\Theta}^R ] ,
\eqno(2.7)
$$
where, in Eq.(2.2),  $q$ and $\ell$ are SU(2)$_L$ doublet fields, and
$f^c$ ($f=u,d,e,\nu$) are SU(2)$_L$ singlet fields. 
The other fields in Eqs.(2.3)-(2.7) have quantum numbers defined in Table 1.

\begin{table}
\begin{center}
\begin{tabular}{|c|ccccc|} \hline
   & 
$H_u$ & $H_d$ &  $E^u$ & $\bar{E}_u$ & $\Theta_{8+1}$\\ \hline
U(3) &  ${\bf 1}$ & ${\bf 1}$ &  ${\bf 6}$ & ${\bf 6}^*$ & 
${\bf 8} + {\bf 1}$   \\
 $R$ & $1$ & $1$ & $1-\bar{r}_E$ & $\bar{r}_E$ & $1$\\ 
Model &  $1$ & $1$ & $3$ & $-2$ & $1$ \\
\hline
\end{tabular}  

\begin{tabular}{|cccccccccc|} \hline
 $\ell$ & $e^c$ & $\nu^c$ & 
$\bar{Y}_e$ & $\bar{\Phi}_0$ & $ E' $ & $X'$ & $\Theta^e$ & 
$\bar{Y}_R$ & ${\Theta}^R$ \\ \hline
${\bf 3}$ & ${\bf 3}$ & ${\bf 3}$ & ${\bf 6}^*$ & ${\bf 6}^*$ &
${\bf 6}$ & ${\bf 6}$ & ${\bf 6}$ &  
 ${\bf 6}^*$ & ${\bf 6}$  \\
$r_\ell$ & $r_e$ & $r_e$ &  $r_{Ye}$ & $r_0$ & $r_{X'}$ & $r_{X'}$ & 
$2-r_{Ye}$ & $r_R$ & $2-r_R$ \\ 
$1$ & $0$ & $0$ & $0$ & $\frac{1}{2}$ & $-1$ & $-1$ & $2$  & 
$2$ & $0$ \\
\hline
\end{tabular}  

\begin{tabular}{|ccccccccccc|} \hline
$q$  & $u^c$ & $d^c$ & $\bar{Y}_u$  & $\bar{\Phi}_u$ & 
${\Theta}^u$ & $\Theta^{u\prime}$ &  $\bar{Y}_d$ & 
$\Theta^d$ & $E$ & $X$  \\ \hline
${\bf 3}$ & ${\bf 3}$ & ${\bf 3}$ & ${\bf 6}^*$ & ${\bf 6}^*$ & ${\bf 6}$ & 
 ${\bf 6}$ & ${\bf 6}^*$ & ${\bf 6}$ & ${\bf 6}$ & ${\bf 6}$ \\
$r_q$ & $r_u$ & $r_d$ &  $r_{Yu}$ &  $r_{Y d} $ &  $2-r_{Yu}$ & 
$2-r_{Y d} $ & $r_{Yd}$ & $2 -r_{Yd}$ & $r_{X}$ & $r_{X}$ 
 \\
$1$ & $-1$ & $+1$ &  $+1$ & $-1$ & $1$ & $3$ & $-1$ & $3$ & $-2$ & $-2$  \\
\hline
\end{tabular}
\end{center}

\begin{quotation}
\caption{
Assignments of $R$ charges, where 
$r_{X'}=r_{Ye}-2 r_0$, $r_{X}=r_{Y d} -2 r_0$ and 
$R(\bar{\Phi} u) = R(\bar{Y}_d) \equiv r_{Yd}$. 
The values in the third raw denote $R$ charge values in 
a special case under the assumptions (A.11) and (A.14).  
For more details, see Eqs.(A.1) - (A.20) in Appendix A. 
}
\end{quotation}
\end{table}


In Eq.(2.6), the third term has been added since it has the same 
$R$ charge as that of $\bar{Y}_u \Theta^u$. 
(See Eq.(A.20) in Appendix A.)
The $\lambda_{0u}$ term plays a role in shifting eigenvalues of the up-quark mass matrix $M_u$ 
by a constant value.
Details of $R$ charge assignments are given in Appendix A.

For the field $E^u$, we assume an additional field $\bar{E}_u$, 
and consider a superpotential with a form
$$
W_E = \lambda_E {\rm Tr}[E^u \bar{E}_u \Theta_{8+1}]
+ \lambda'_E {\rm Tr}[E^u \bar{E}_u]\, {\rm Tr}[\Theta_{8+1}]  ,
\eqno(2.8)
$$
where $\Theta_{8+1}$ is a field  ${\bf 8}+{\bf 1}$ of
U(3) with $\langle \Theta_{8+1} \rangle =0$.
The superpotential $W_E$ leads to 
$\langle E^u \rangle \langle \bar{E}_u\rangle \propto {\bf 1}$. 
We assume that the form 
$$
\langle \bar{E}_u \rangle \propto \langle E^u \rangle = 
v_{E} \,{\rm diag}(1,1,1) ,
\eqno(2.9)
$$ 
is given by a specific form of the solutions 
$\langle E^u \rangle \langle \bar{E}_u\rangle \propto {\bf 1}$.

In this paper, we do not discuss a superpotential which gives 
the observed charged lepton mass spectrum. 
We only use the observed charged lepton mass 
values as input values in $\langle \bar{Y}_e \rangle_e$.

Under the assumption that all $\Theta$ fields take 
$\langle \Theta \rangle =0$, SUSY vacuum conditions  
lead to VEV relations\footnote{
For example, in obtaining the relation (2.10), 
we have assumed a vacuum with $\langle \Theta^e \rangle =0$, 
so that the  conditions $\partial W/\partial Y_e =0$ and 
$\partial W/\partial \Phi_0 =0$ do not affect other VEV relations
obtained from SUSY vacuum conditions 
$\partial W/\partial \Theta_A =0$ ($A\neq e$).
We assume that the observed SUSY symmetry breaking is induced by
a gauge mediation mechanism (not including family symmetry), 
so that our VEV relations among yukawaons are still valid 
in the quark and lepton sectors after the SUSY is broken .
}
 from  Eqs.(2.3)-(2.7).
That is, instead of Eqs.(1.2) - (1.8) in the previous model,  
we obtain the following mass matrix relations: 
$$
M_e \propto \langle \bar{Y}_e \rangle \propto 
\langle \bar{\Phi}_0 \rangle ({\bf 1} + a_e S_2 ) 
\langle \bar{\Phi}_0 \rangle , 
\eqno(2.10)
$$
$$
M_d \propto \langle \bar{Y}_d \rangle \propto 
\langle \bar{\Phi}_0 \rangle ({\bf 1} + a_d e^{i\alpha_d} S_3 ) 
\langle \bar{\Phi}_0 \rangle , 
\eqno(2.11)
$$
$$
\langle \bar{\Phi}_u \rangle \propto 
\langle \bar{\Phi}_0 \rangle ({\bf 1} + a_u e^{i\alpha_u} S_3 ) 
\langle \bar{\Phi}_0 \rangle  , 
\eqno(2.12)
$$
$$
M_u \propto \langle \bar{Y}_u \rangle 
\propto \langle \bar{\Phi}_u \rangle \cdot {\bf 1} 
\cdot \langle \bar{\Phi}_u \rangle +(v_{\Phi u})^2 \zeta_u {\bf 1} , 
\eqno(2.13)
$$
$$
M_R \propto \langle \bar{Y}_R \rangle \propto 
\langle \bar{\Phi}_u \rangle\cdot {\bf 1} 
\cdot  \langle \bar{Y}_e \rangle 
+ \langle \bar{Y}_e \rangle\cdot {\bf 1} 
\cdot  \langle \bar{\Phi}_u \rangle . 
\eqno(2.14)
$$
Here, we can take a diagonal basis of 
$\langle \bar{\Phi}_0 \rangle$ without loosing the generality: 
$$
\langle \bar{\Phi}_0 \rangle = {\rm diag}(v_1, v_2, v_3)
= v_0 \, {\rm diag}(x_1, x_2, x_3) ,
\eqno(2.15)
$$ 
where we have normalized $x_i$ as $x_1^2+x_2^2+x_3^2=1$.
The neutrino mass matrix $M_\nu$ is given by a seesaw type 
$M_\nu = m_D M_R^{-1} m_D^T$ with $m_D \propto M_e$ similar
to Eq.(1.4) [but there is no $m_{0\nu}^{-1}$]. 
Note that the previous relations (1.2) - (1.8) were given at a diagonal
basis of the VEV $\langle \Phi_e \rangle$, while present relations 
(2.10) - (2.14) are given at a diagonal basis of the VEV 
$\langle \bar{\Phi}_0 \rangle$. 
Here the numerical matrices $S_3$ and $S_2$ are 
defined by
$$
S_3 = \frac{1}{3}  \left(
\begin{array}{ccc}
1 & 1 & 1 \\
1 & 1 & 1 \\
1 & 1 & 1
\end{array} \right) , \ \ \ \ 
S_2 = \frac{1}{2}  \left(
\begin{array}{ccc}
1 & 1 & 0 \\
1 & 1 & 0 \\
0 & 0 & 0
\end{array} \right) ,
\eqno(2.16)
$$
at the diagonal basis of the VEV 
$\langle \bar{\Phi}_0 \rangle$.
Note that the VEV matrix $\langle \bar{Y}_e \rangle$ in Eq.(2.10) is 
no more diagonal in this basis.

In obtaining the mixing matrices, the common coefficients are not important.
Here we have taken 
 $v_{E'}=v_{X2}$ and $v_{E}=v_{X3}$ for simplicity. 
The $\zeta_u$ term in Eq.(2.13) comes from the new term given in Eq.(A.20).   
This term contributes to the up-quark mass ratios, while not
to the up-quark mixing matrix, so that it does not 
change the predictions for the neutrino mixing parameters.
We suppose that the contribution from such the higher dimensional term 
(A.20) is considerably small, so that it also does not visibly 
affect the up-quark mass ratio $m_c/m_t$, although it can slightly 
affect $m_u/m_c$.

\section{Numerical results in the up-quark and neutrino mass matrices}

In this section, we investigate whether the new VEV matrix relations (2.10) -
(2.14) can well describe the observed neutrino mixing parameters 
together with the observed up-quark mass ratios or not.

Since the charged lepton mass matrix given by Eq.(2.10) is 
not diagonal,  
the lepton mixing matrix [Pontecorvo-Maki-Nakagawa-Sakata (PMNS) \cite{PMNS}
mixing matrix] $U$ in the present conventions is defined by 
$$
U=U_{eL}^\dagger U_{\nu L} ,
\eqno(3.1)
$$
where $U_{eL}$ and $U_{\nu L}$ are defined by
$$
U_{eL}^T \langle \bar{Y}_e \rangle U_{eL} = 
\langle \bar{Y}_e \rangle^{diag} \equiv 
\langle \bar{Y}_e \rangle_e ,
\eqno(3.2)
$$
$$
U_{\nu L}^\dagger (M_\nu^\dagger M_\nu) U_{\nu L} = 
(M_\nu^\dagger M_\nu)^{diag} ,
\eqno(3.3)
$$
and $M_\nu$ is given by
$$
M_\nu = \frac{y_\nu^2}{\lambda_R} \left(
\frac{\langle H_u^0 \rangle }{\Lambda} \right)^2 
\langle \bar{Y}_e \rangle \langle \bar{Y}_R \rangle^{-1}
\langle \bar{Y}_e \rangle .
\eqno(3.4)
$$
Neutrino mixing parameters we discuss are 
$\tan^2 \theta_{solar}=|U_{12}|^2/|U_{11}|^2$, 
$\sin^2 2\theta_{atm}=4|U_{23}|^2 |U_{33}|^2$, and $|U_{13}|^2$. 
Here $U_{ij}$ are the matrix elements of the lepton mixing matrix 
defined by (3.1). 

The matrix $M_u$ in (2.13) is diagonalized as 
$$
U_{uL}^\dagger (M_u^\dagger M_u) U_{uL} 
= (M_u^\dagger M_u)^{diag} .
\eqno(3.5)
$$
Here $U_{uL}$ is a mixing matrix among left-handed up-quarks $u_{Li}$ . 
(In the present paper, the mass matrices (i.e. 
$\langle \bar{Y}_f\rangle$) are defined by Eq.(2.2).
Therefore, the conventions of the mixing matrices 
are somewhat changed from the conventional ones.)
Note that since the VEV matrix $\langle \bar{\Phi}_u \rangle$ is
complex and $\langle \bar{Y}_u \rangle$ is given by Eq.(2.13), 
the diagonalization of the up-quark mass matrix must be done by 
Eq.(3.5). 

\vspace{2mm}

\noindent
{\bf 3.1 \ Parameters in the model}
 
The mass matrices for quarks and neutrinos in the O(3) model
have been described in terms of the fundamental VEV matrix 
$\langle \Phi_e\rangle$. On the other hand, the fundamental VEV matrix 
in the present model 
is $\langle \bar{\Phi}_0\rangle$ defined by Eq.(2.10) 
in which we have new parameter $a_e$. 
Thus the number of parameters 
are increased by one compared with the previous model (1.2).
On the other hand, we cannot bring neither 
the $\xi_\nu$ term given in Eq.(1.5) nor 
$\langle P_u \rangle_u$ defined in Eq.(1.10)
into the present model, so that there are no parameters which
are corresponding to $\xi_\nu$ and $P_u$. 

The VEV of $\langle \bar{\Phi}_0\rangle={\rm diag}(v_1,v_2,v_3)$ 
is related to the charged lepton mass matrix $M_e$ 
as follows:
$$
M_e = k \left(
\begin{array}{ccc}
\left(1 +\frac{1}{2}a_e\right) v_1^2 & \frac{1}{2}a_e v_1 v_2
& 0 \\
\frac{1}{2}a_e v_1 v_2 & \left(1 +\frac{1}{2}a_e\right) v_2^2
& 0 \\
0 & 0 & v_3^2 
\end{array} \right) ,
\eqno(3.6)
$$
where $k=-(\lambda_e/\mu_e \Lambda)(y_e \langle H^0_d\rangle
/\Lambda)v_{X'}$, 
so that we obtain
$$
m_e+m_\mu = k \left(1 +\frac{1}{2}a_e\right)(v_1^2+v_2^2) ,
\eqno(3.7)
$$
$$
m_e m_\mu = k^2 (1+a_e) v_1^2 v_2^2 ,
\eqno(3.8)
$$
and $m_\tau = k v_3^2$. 
Here, since we are interested only in the relative ratios
among the eigenvalues of the charged lepton mass matrix
$M_e$, the common coefficient $k$ is not a parameter of 
the model.
The 3 parameters $a_e$, $v_1/v_2$ and $v_2/v_3$ are
sufficient to determine the two charged lepton mass ratios
and charged lepton mixing matrix $U_e$ which is described 
only by one parameter $\theta_{12}^e$.
[The mixing angle $\theta^e_{12}$ is not observable.
The observed quantities are parameters of the lepton mixing
matrix defined by Eq.(3.1).]
Therefore, when we give a value of the parameter $a_e$, 
the values of $v_i$ are completely determined by
the input values of the charged lepton masses. 
In other words, even when we give three charged lepton 
masses as the inputs, one of the free parameters still remains.

Thus, in the present model, we have 4 parameters $a_e$, 
$a_u$, $\alpha_u$ and $\zeta_u$ (except for the input values 
$m_{e}$, $m_{\mu}$, and $m_{\tau}$) 
for the up-quark and neutrino mass matrices. 
On the other hand, 
the number of the predictable quantities are 12, i.e., 
2+2+2 mass ratios (up-quark, charged lepton and neutrino 
mass ratios) and 4+2 PMNS mixing parameters (including 
two Majorana phases). 
At present, we know 6 observed values of $\sqrt{m_u/m_c}$,
$\sqrt{m_c/m_t}$, $\tan^2 \theta_{solar}$, $\sin^2 2 \theta_{atm}$,
$\sin^2 2 \theta_{13}$, and $R_\nu = \Delta m^2_{21}/\Delta m^2_{32}$
in addition to the charged lepton masses. 

The term $\zeta_u {\bf 1}$ in $M_u$ given in Eq.(2.13) does not affect 
the up-quark mixing
matrix, so that it also affects neither quark or lepton mixing
matrices.
Since we suppose $|\zeta_u|^2 \ll 1$, the term almost does  not
affect $\sqrt{m_c/m_t}$, although it can slightly affect $\sqrt{m_u/m_c}$.
As a result, the present model predicts 11 observables 
by using the three parameters $(a_e, a_u, \alpha_u)$.
In other words, the value of $\sqrt{m_u/m_c}$ is not `` prediction",
and it is a quantity which can be adjustable by the additional parameter
$\zeta_u$ freely.

\vspace{2mm}

\noindent
{\bf 3.2 \ Numerical results}

Now let us show the results of numerical analysis of the model.
First, we show, in Fig.~1, the $a_u$ dependences of the quantities 
$\sqrt{m_c/m_t}$, 
$\tan^2 \theta_{solar}$, and $\sin^2 2\theta_{atm}$ 
with taking typical values of  
$a_e=3,\ 30,\ 100$ and $\alpha_u =0^\circ,\ 15^\circ$ 
in order to see rough parameter behaviors. 

\begin{figure}[t!]
  \includegraphics[width=50mm,clip]{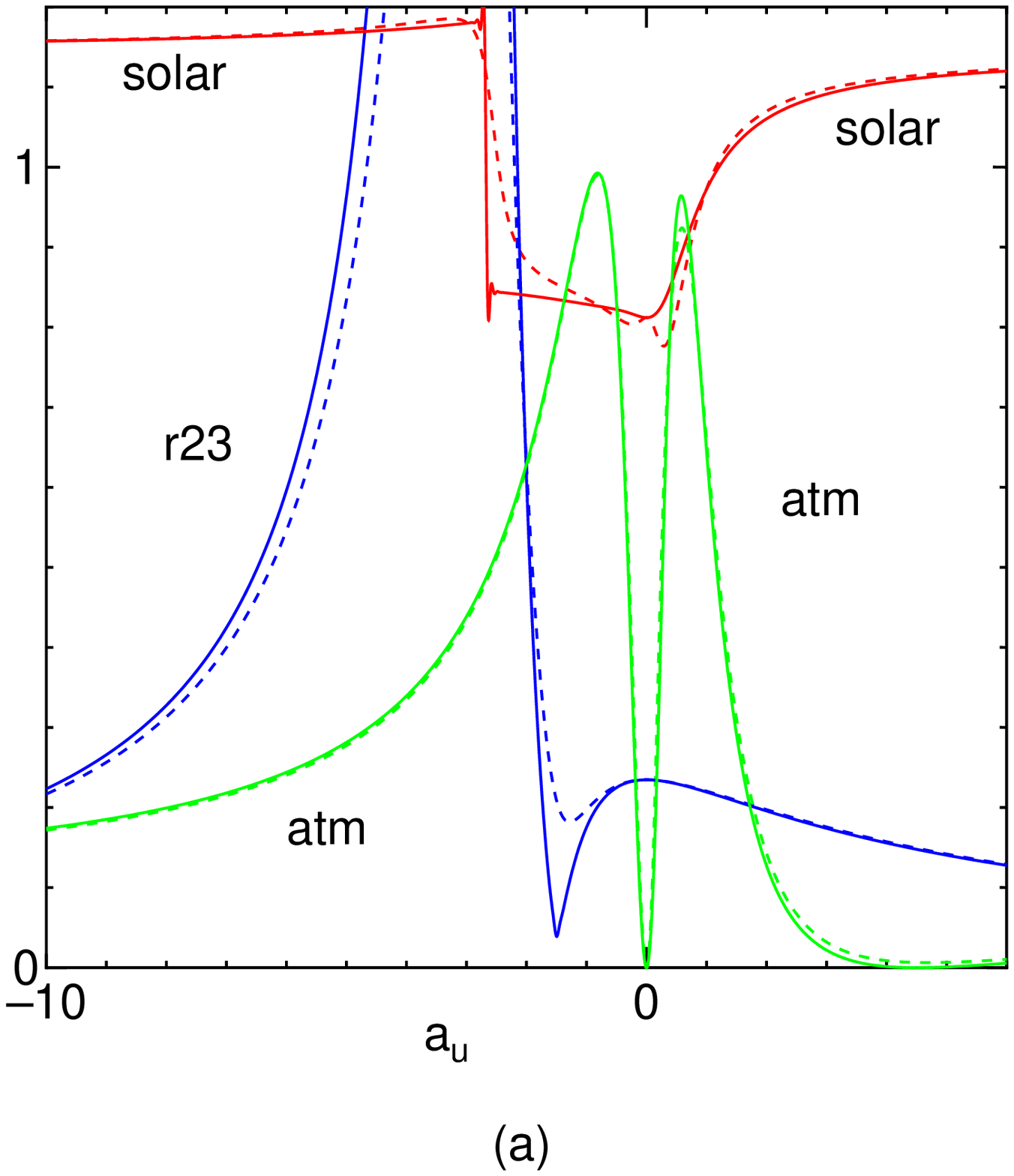}
  \includegraphics[width=50mm,clip]{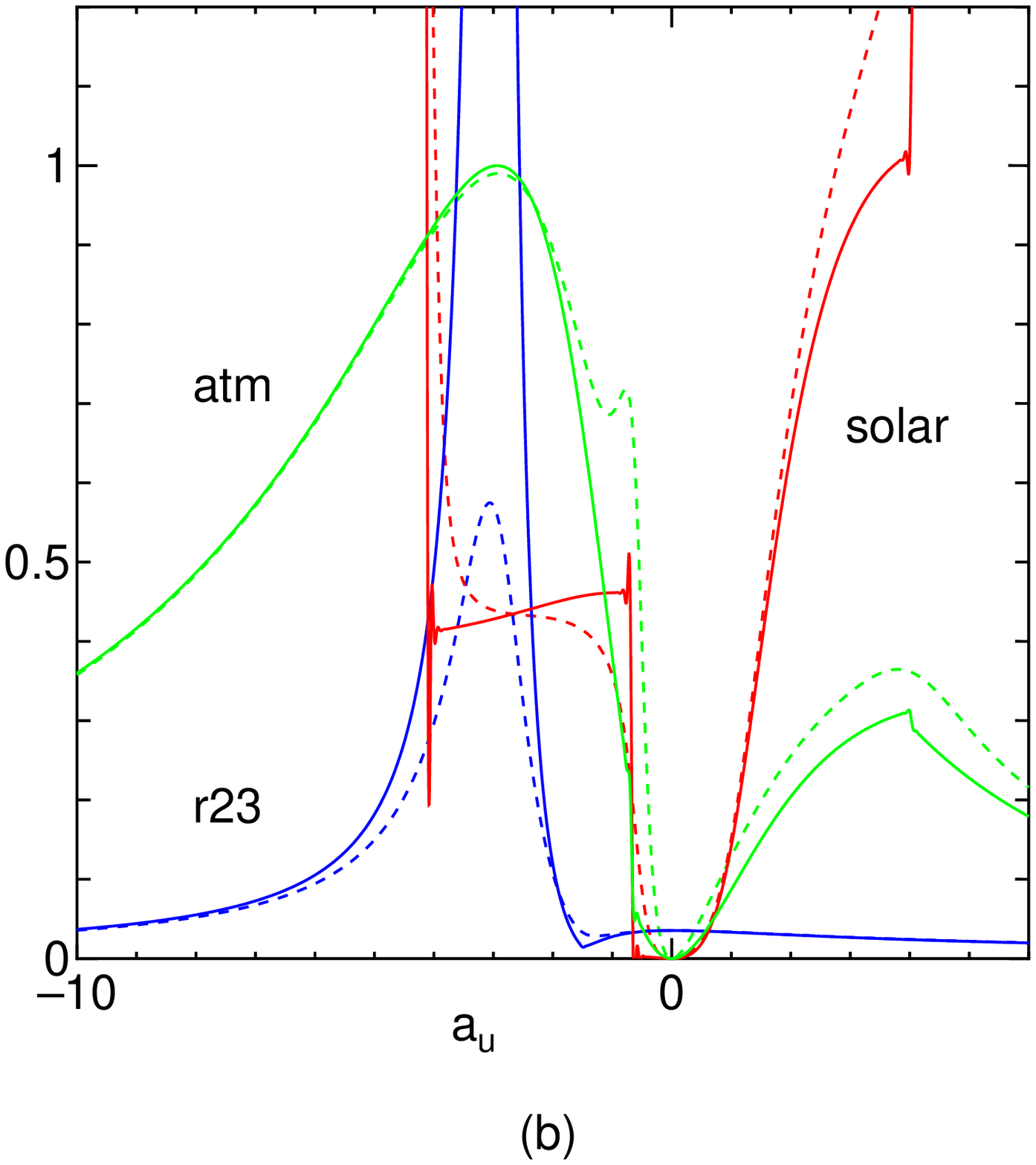}  
  \includegraphics[width=50mm,clip]{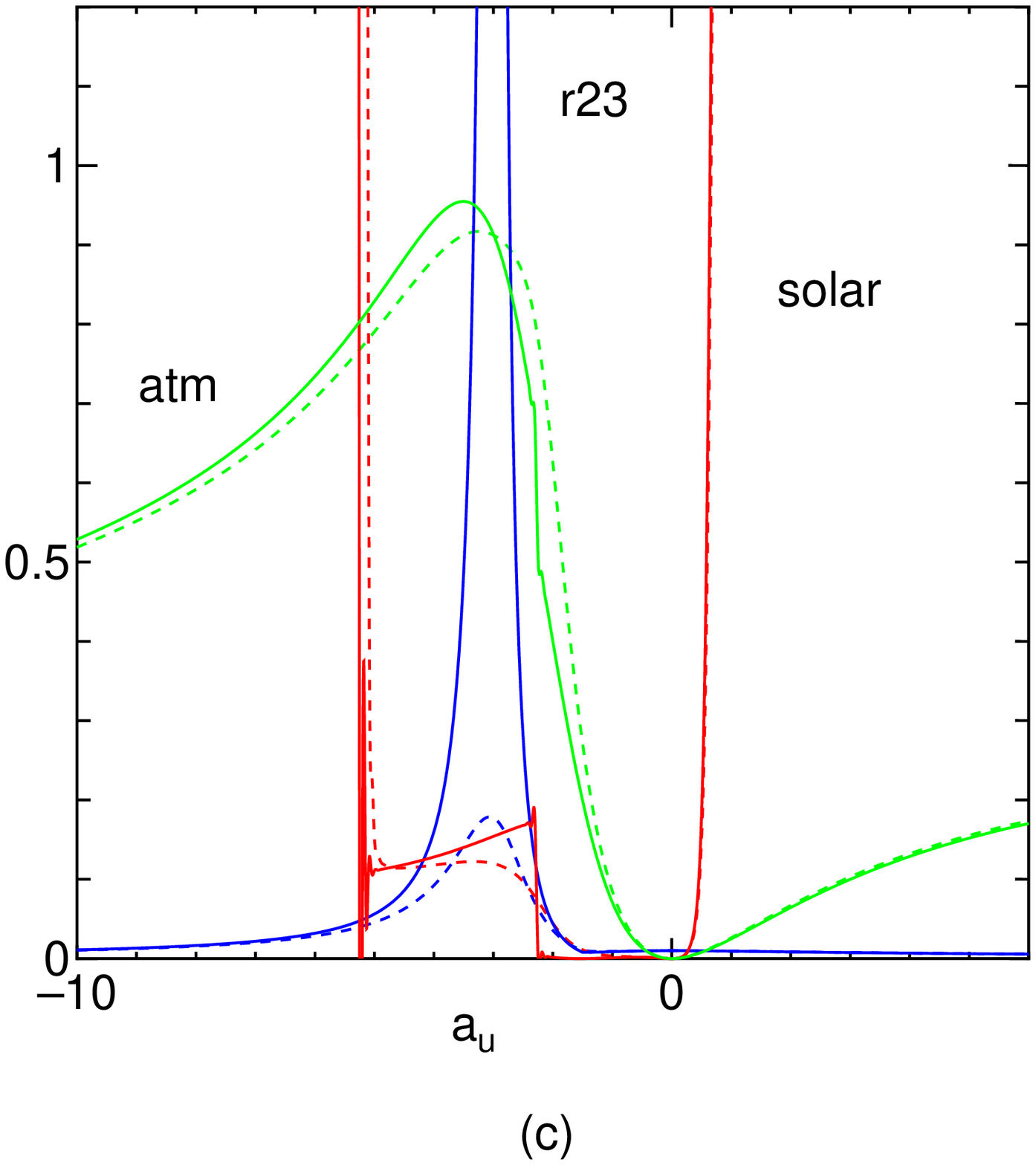}   
  \caption{$\sqrt{m_c/m_t}$, $\tan^2\theta_{solar}$, and 
$\sin^2 2\theta_{atm}$ versus a parameter $a_u$
for typical parameter values $a_e=3$ (Fig.~1 (a)), 
$a_e=30$ (Fig.~1 (b)), and $a_e=100$ (Fig.~1 (c)) 
with $\alpha_u=0^\circ$ (solid curves) and
$\alpha_u = 15^\circ$ (dashed curves).
Curves ``r23", ``solar", and  ``atm" denote 
``r23"$=\sqrt{m_c/m_t}\times 10$, ``solar"$=\tan^2\theta_{solar}$, 
and ``atm"$=\sin^2 2\theta_{atm}$, respectively.
}
  \label{au-depnd}
\end{figure}

As seen in Fig.~1, we can find that 
(i) the value of $\sqrt{m_c/m_t}$ takes a maximum value 
at $a_u \sim -3$ insensitively to the values of 
$a_e$ and $\alpha_u$; 
(ii) since the maximum value of $\sin^2 2\theta_{atm}$
shows $\sin^2 2\theta_{atm} \simeq 1$ which is in favor of the 
observed value, we must search for a parameter set 
$(a_e, a_u, \alpha_u)$ which gives 
a maximum value of $\sin^2 2\theta_{atm}$; 
(iii) a case with a small value of $a_e$ gives a large value of 
$\tan^2 \theta_{solar}$ compared with the observed value 
$\tan^2 \theta_{solar} \sim 0.5$ (see Fig.~1 (a)), so that 
such a case is ruled out; on the other hand, a case with 
a large value of $a_e$ gives a small $\tan^2 \theta_{solar}$
(see Fig.~1 (c)), so that such a case is also ruled out;
(iv) as a result, a region of $(a_e, a_u)$ which can give  
$\sin^2 2\theta_{atm} \simeq 1$ and 
$\tan^2 \theta_{solar} \sim 0.5$ is $(a_e, a_u) \sim 
(30, -3)$.

\begin{figure}[t!]
  \includegraphics[width=50mm,clip]{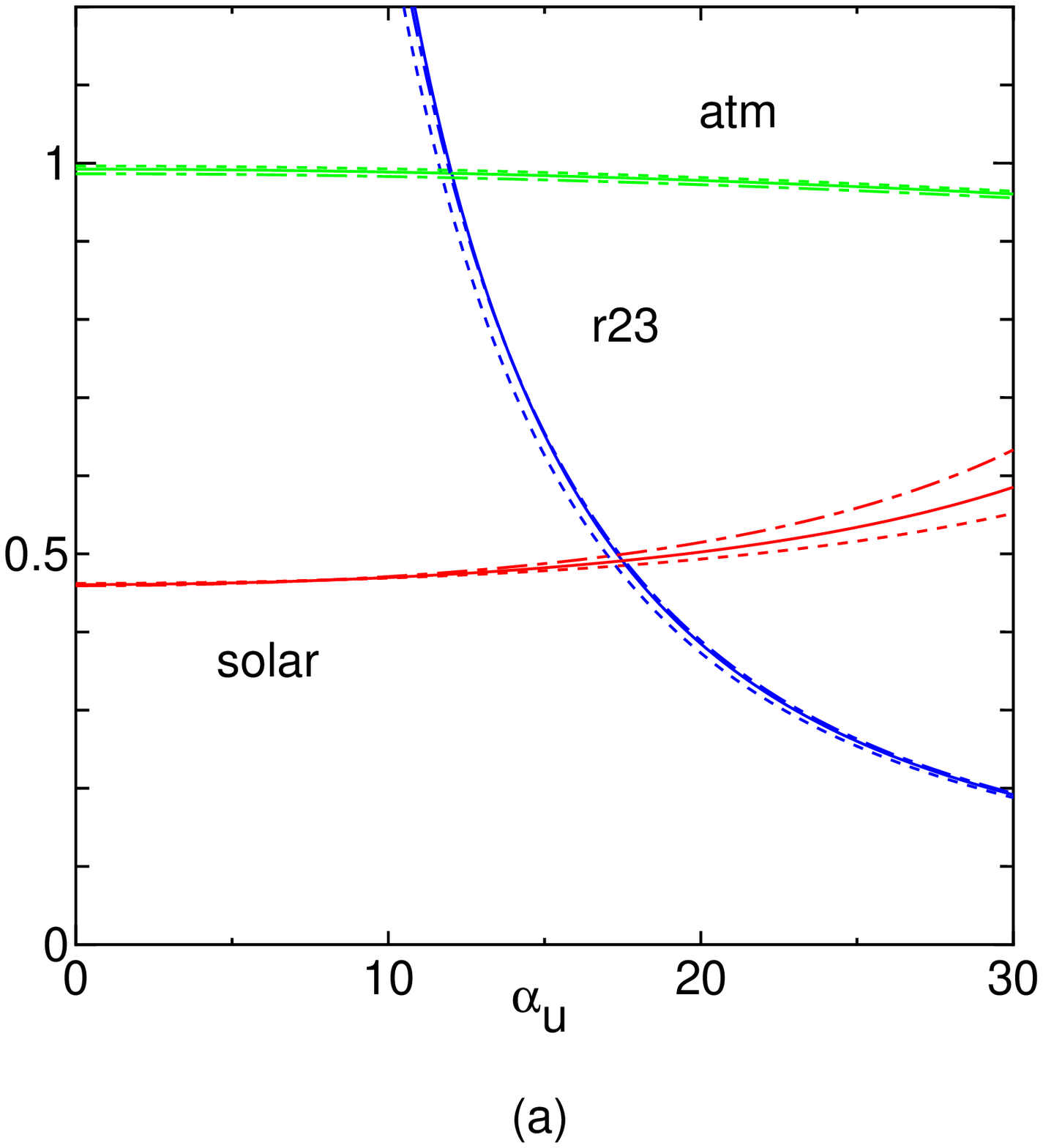}
  \includegraphics[width=50mm,clip]{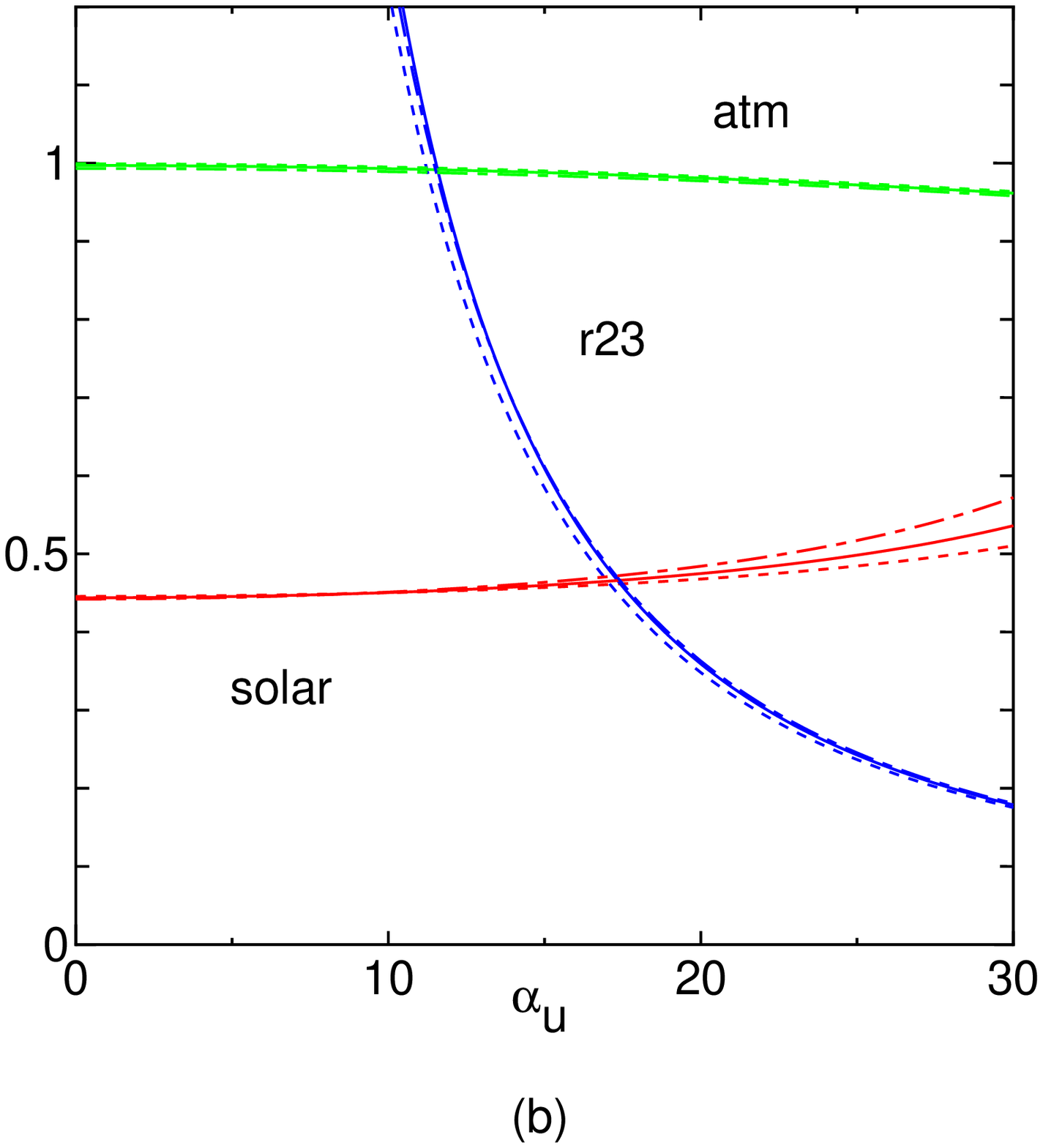}  
  \includegraphics[width=50mm,clip]{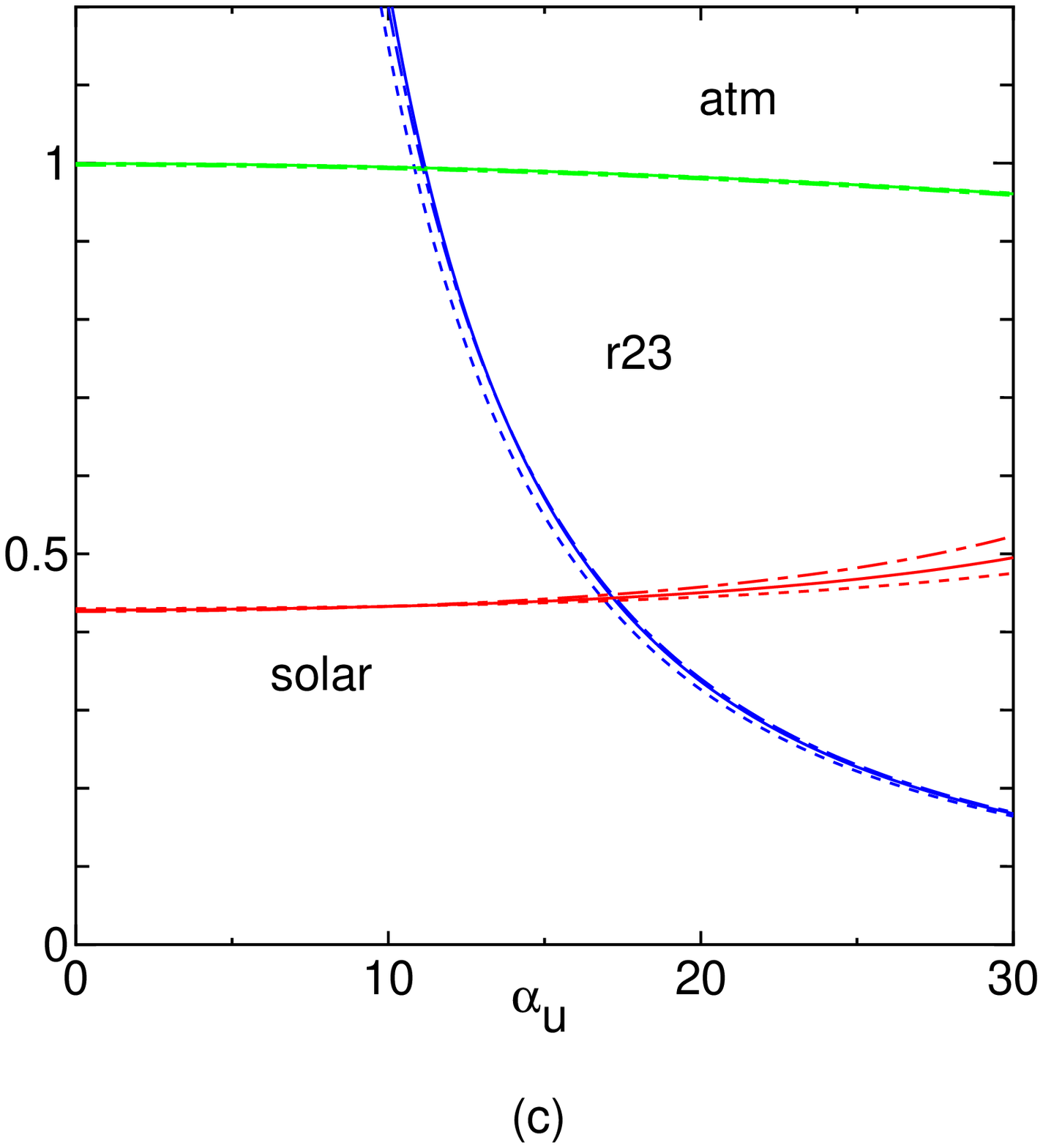}   
  \caption{$\sqrt{m_c/m_t}$, $\tan^2\theta_{solar}$, and 
$\sin^2 2\theta_{atm}$ versus a parameter $\alpha_u$
for typical parameter values $a_e=26$ (Fig.~2 (a)), 
$a_e=28$ (Fig.~2 (b)), and $a_e=30$ (Fig.~2 (c)) 
with $a_u = -2.9$ (dashed curves), $a_u=-3.0$ (solid curves), and 
$a_u = -3.1$ (dot-dashed curves).
Curves ``r23", ``solar", and  ``atm" denote 
``r23"$=\sqrt{m_c/m_t}\times 10$, ``solar"$=\tan^2\theta_{solar}$, 
and ``atm"$=\sin^2 2\theta_{atm}$, respectively.
}
  \label{alpu-depnd}
\end{figure}

\begin{figure}[t!]
  \includegraphics[width=60mm,clip]{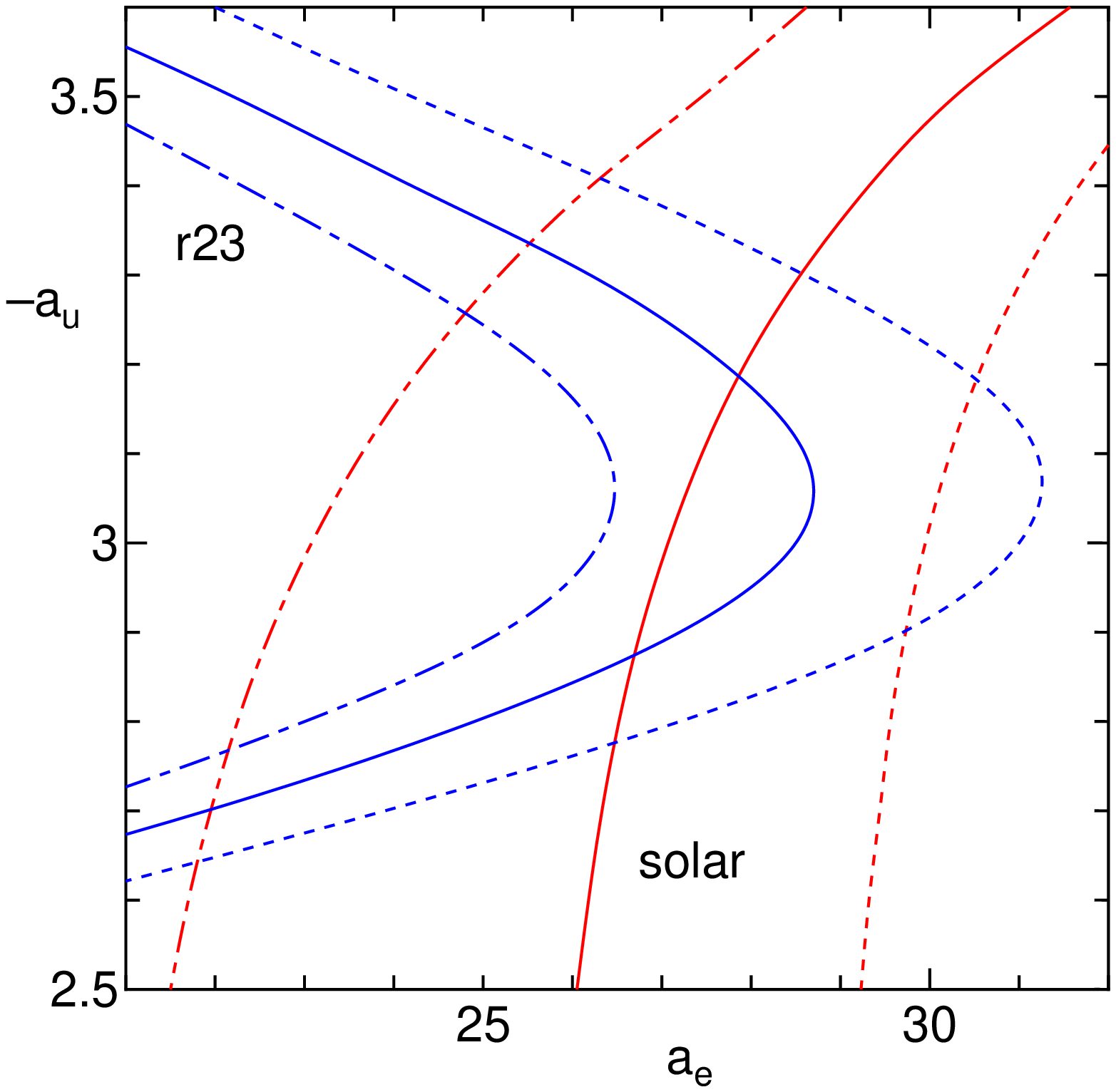}
  \caption{Contour lines of $\sqrt{m_c/m_t}$ and $\tan^2\theta_{solar}$ 
in the $(a_e, a_u)$ plane in the case of  $\alpha_u=15^\circ$.
As input values, $\sqrt{m_c/m_t}=0.0600^{+0.0045}_{-0.0047}$
and $\tan^2\theta_{solar}^{obs}=0.47^{+0.05}_{-0.03}$ 
have been used.
Curves with dot-dash, solid, and dash denote the 
upper, center, and lower observed values, respectively. 
Curves ``r23" and ``solar" denote 
``r23"$=\sqrt{m_c/m_t}\times 10$ and  ``solar"$=\tan^2\theta_{solar}$, respectively.
}
  \label{ae-au}
\end{figure}

Next, in order to determine parameter values $(a_e, a_u, \alpha_u)$,
let us illustrate, in Fig.~2, the $\alpha_u$ behaviors of 
$\sqrt{m_c/m_t}$, $\tan^2 \theta_{solar}$ and 
$\sin^2 2\theta_{atm}$ at $a_e \sim 28$ and $a_u \sim -3$.
From Fig.~2, we search for the value $\alpha_u$ which gives 
the observed value \cite{q-mass} 
$\sqrt{m_c/m_t}=0.0600^{+0.0045}_{-0.0047}$ at $\mu=m_Z$.
We find that the value $\alpha_u \simeq 15^\circ$ can give a reasonable 
fit $\sqrt{m_c/m_t}=0.0600$ insensitively to the other parameters.

Therefore, by fixing the value $\alpha_u = 15^\circ$, 
we illustrate the contour lines of $\sqrt{m_c/m_t}$ and 
$\tan^2 \theta_{solar}$ in the $(a_e, a_u)$ plane  in Fig.~3.
The curves denote $(a_e, a_u)$ which gives the observed values
 $\sqrt{m_c/m_t}=0.0600^{+0.0045}_{-0.0047}$ \cite{q-mass} 
and $\tan^2\theta_{solar}^{obs}=0.47^{+0.05}_{-0.03}$ 
\cite{PDG10}.
As seen in Fig.~3, we have two intersection points of the 
curves of $\sqrt{m_c/m_t}$ and $\tan^2\theta_{solar}$.
For the center values $\sqrt{m_c/m_t}=0.060$ and 
$\tan^2\theta_{solar}=0.47$, the solutions $(a_e, a_u, \alpha_u)$
are 
$$
(26.7, -2.88, 15^\circ), \ \ \ \ (27.9, -3.18, 15^\circ) .
\eqno(3.9)
$$
We list our prediction values for these parameter solutions
in Table 1. 
Of the two solutions obtained from the input data $\sqrt{m_c/m_t}$ 
and $\tan^2\theta_{solar}$, Table 2 suggests that we 
should take the former one considering the observed 
value of $R_\nu$ \cite{PDG10}
$$
R_\nu \equiv \frac{\Delta m^2_{solar}}{\Delta m^2_{atm}} 
= (3.12^{+0.27}_{-0.23}) \times 10^{-2} .
\eqno(3.10)
$$
For reference, we also illustrate the behavior of predicted values for
input values $(a_e, a_u, \alpha_u)$ around the parameter solutions 
(3.9) in Fig.~4.
As seen in Fig.~4, the predicted values $\sin^2 2\theta_{atm}$ 
and $\tan^2 \theta_{solar}$ are insensitive to the 
parameter values $a_e$, $a_u$, and $\alpha_u$ around the
values $(a_e, a_u, \alpha_u)=(26.7, -2.88,15^\circ)$.
However, $\sqrt{m_c/m_t}$ and $|U_{13}|^2$ (and also 
$R_\nu$) are somewhat dependent on these parameters.
Since these parameter values are mainly obtained by taking 
the input value $\sqrt{m_c/m_t}=0.0600$, if the input value
changes, then the predicted values will also change.

\begin{table}
\begin{center}
\begin{tabular}{|c|ccccc|} \hline
Input $(a_e, a_u, \alpha_u)$ & $\sqrt{m_c/m_t}$ & $\tan^2\theta_{solar}$
& $\sin^2 2\theta_{atm}$ & $\sin^2 2\theta_{13}$ & $R_\nu$ \\ \hline
$(26.7, -2.88, 15^\circ)$ & $0.0603$ & $0.470$ & $0.990$ & 
$0.015$ & $0.0303$ \\
$(27.9, -3.18, 15^\circ)$ & $0.0601$ & $0.469$ & $0.980$ & 
$0.011$ & $0.0204$ \\ \hline
\ \ \ \ upper & \ {\footnotesize $ +0.0045$} & \ {\footnotesize $+0.05$} &  
&  & \ {\footnotesize $+0.0027$}  \\
Observed value & $0.0600$ & $0.47$ & $>0.92$ \cite{PDG10} & $<0.15$  \cite{PDG10}& $0.0312$  \\
\ \ \ \ lower & \ {\footnotesize $-0.0047$} & \ {\footnotesize $-0.03$} &
  &  & \ {\footnotesize $-0.0023$} \\ \hline
\end{tabular}

\end{center}
\begin{quotation}
\caption{
Predicted values for the parameter values $(a_e, a_u, \alpha_u)$. 
}
\end{quotation}
\end{table}

%

\begin{figure}[t!]
  \includegraphics[width=60mm,clip]{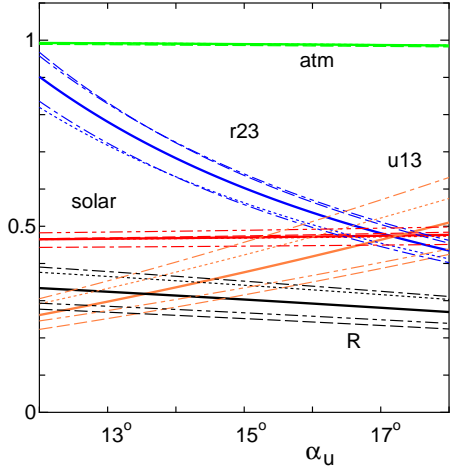}
  \caption{ Predicted values versus a parameter $\alpha_u$
for $a_e=26.7$ and $a_u=-2.88$. 
Curves ``r23", ``solar", ``atm", ``u13", and ``R" denote 
``r23"$=\sqrt{m_c/m_t}\times 10$, ``solar"$=\tan^2\theta_{solar}$, 
``atm"$=\sin^2 2\theta_{atm}$, ``u13"$=|U_{13}|^2 \times 100$, 
and ``R"$=R_\nu \times 10$, respectively. 
For reference, curves for $(a_e, a_u)=(26.7, -3.00)$ 
(dash curve),  $(26.7, -2.80)$ (dot curve), $(29,-2.88)$ 
(dot dash curve), and $(25, -2.88)$ (2-dot dash curve) are
illustrated in addition to the curve (solid) for $(26.7, -2.88)$.
}
  \label{all}
\end{figure}

So far, we have not discussed the value of $m_u/m_c$.
In the present model, the value of $m_u/m_c$ is always adjustable 
by the parameter $\zeta_u$ given in 
Eq.(2.13) without affecting other predicted values.
In order to fit the predicted value of $\sqrt{m_u/m_c}$ to the 
observed value \cite{q-mass} $\sqrt{m_u/m_c}=0.0453^{+0.012}_{-0.010}$, 
we choose $\zeta_u$ as $\zeta_u= 3.8 \times 10^{-7}$. 
As seen in Table 3, the value of $\zeta_u$ almost does not change the 
numerical predictions given in Table 2.  

In conclusion, we take the parameter set 
$$
(a_e, a_u, \alpha_u, \zeta_u)= (26.7, -2.88, 15^o, 3.8\times 10^{-7}) .
\eqno(3.11)
$$
Then, we predict neutrino masses
$$
m_{\nu 1} \simeq 0.012\ {\rm eV}, \ \ m_{\nu 2} \simeq 0.015 \ {\rm eV}, 
\ \ m_{\nu 3} \simeq 0.051 \ {\rm eV}  ,
\eqno(3.12)
$$
by using the input value $\Delta m^2_{32}\simeq 0.0024$ eV$^2$.
We also predict the effective Majorana mass $\langle m_{ee}\rangle$ 
in the neutrinoless double beta decay \cite{Doi1981} 
$$
\langle m_{ee}\rangle =\left|m_1 U_{e1}^2 +m_2 U_{e2}^2 
+m_3 U_{e3}^2\right| \simeq 0.0039 \ {\rm eV}.
\eqno(3.13)
$$
It is worthwhile noticing approximately degenerate neutrino masses
$m_{\nu 1} \sim m_{\nu 2}$.

\begin{table}
\begin{center}
\begin{tabular}{|c|ccccccc|} \hline
Input $\zeta_u$ & $\sqrt{m_u/m_c}$ & $\sqrt{m_c/m_t}$ & $\tan^2\theta_{solar}$
& $\sin^2 2\theta_{atm}$ & $\sin^2 2\theta_{13}$ &
$\delta_{CP}$ ($J$) &  $R_\nu$ \\ \hline
$0$ & $0.0180$ & $0.0603$ & $0.470$ & $0.990$ & $0.015$ & 
$-102.5^\circ$ ($-0.0139$) & $0.0303$ \\
$3.8 \times 10^{-7}$ & $0.0454$ & $0.0602$ & $0.470$ & $0.990$ & 
$0.015$ & $-102.5^\circ$ ($-0.0139$)  & $0.0303$ \\ \hline
\end{tabular}

\end{center}
\begin{quotation}
\caption{
Predicted values versus $\zeta_u$ parameter.  
Other parameters $(a_e, a_u, \alpha_u)$ have 
taken the same values $(26.7, -2.88, 15^\circ)$ as those in Table 2. 
}
\end{quotation}
\end{table}


\vspace{3mm}


\section{Concluding remarks}

In this paper we have found out a special form of the neutrino
mass matrix $M_\nu$ based on a yukawaon model with U(3) family symmetry, 
which has quite few free parameters. With this form of $M_\nu$,  
the $M_\nu$ can give reasonable predictions in spite 
of having no adjustable parameters, i.e. the $M_\nu$ is simply given 
by the form (1.15), and the mass matrices $M_e$ and $M_u$ 
include parameters $a_e$ and $a_u$, respectively, which are fixed by their observed mass ratios.
In this yukawaon model, the yukawaon VEV matrices are described 
in terms of a new fundamental VEV matrix $\langle \bar{\Phi}_0\rangle$. 
For example, the yukawaon VEV matrix $\langle \bar{Y}_e \rangle$ 
for the charged leptons is given by (2.10) which has 
the structure of $({\bf 1}+a_e S_2)$  
with a new parameter $a_e$. 
This structure in $\langle \bar{Y}_e \rangle$ has been chosen 
from a phenomenological point of view and there is no reason 
why $\langle \bar{Y}_e \rangle$ takes such a form. 
Nevertheless if we accept the form (2.10), 
then we can obtain a simple form of VEV matrix $\langle \bar{Y}_R \rangle$ 
for right-handed neutrinos without introducing the somewhat 
strange VEV matrix $P_u$ and $\xi_\nu$ term 
that were introduced in the O(3) model  
to get the observed nearly tribimaximal neutrino mixing
\cite{tribi}. 

The new model has only four parameters $(a_e, a_u, \alpha_u, \zeta_u)$
as far as the up-quark and lepton sectors are concerned. 
on the other hand, we have 12 observable quantities 
(2 up-quark mass ratios, 
4 lepton mass ratios, and 4+2 lepton mixing parameters). 
The parameter $\zeta_u$ affects only the prediction of $m_u/m_c$, 
so that we have fixed it by the observed value of $r^u_{12}=\sqrt{m_u/m_c}$.
The parameter $\alpha_u$ is sensitive only to $m_c/m_t$, so that
we have fixed by the observed value of $r^u_{23}=\sqrt{m_c/m_t}$
as seen in Fig.~2 (b).
The parameter $a_e$ is determined from the cross point of the 
predicted values of $r^u_{23}$ and $\tan^2 \theta_{solar}$ 
in the $a_e$-$a_u$ plane.  
Note that we have used only the observed values of $r^u_{23}$ and 
$\tan^2 \theta_{solar}$ in order to fix the three parameters
$(a_e, a_u, \alpha_u)$.
Although we have tacitly used $\sin^2 2\theta_{atm} \sim 1$, 
we have not used the observed value of $\sin^2 2\theta_{atm}$ explicitly.  

On the other hand, for the remaining 2 down-quark mass ratios and 4 CKM 
mixing parameters, we have additional 2 parameters ($a_d$ and $\alpha_d$). 
Regrettably, we cannot obtain reasonable predictions with the two parameters,
although we can fit the values of down-quark mass ratios and $V_{us}$.
The situation is the same as in the previous O(3) model.
We must introduce a phase matrix $P_d$ with two parameters $(\phi_1, \phi_2)$
and a common mass shift term $m_{0d} {\bf 1}$.
Then, five parameters can fit six observables barely. 
Therefore, the model is not so attractive for down-quark sector.
In this paper, we did not demonstrate the explicit numerical fitting 
for down-quark mass rations and CKM mixing parameters. 

The present U(3) model have the following interesting features in the lepton sector:

\noindent
(i) The model predicts $\sin^2 2\theta_{13} \sim 0.015$. 
In the previous O(3) model, the predicted value of 
$|U_{13}|^2$ was invisibly small, i.e. $|U_{13}|^2 \sim 10^{-4}$.
The T2K experiment \cite{T2K11} put a constraint 
$ 0.03 < \sin^2 2\theta_{13} <0.28$  ($90\%$ C.L.) for 
$\delta_{CP}=0$ and a normal hierarchy. 
Our predicted value $\sin^2 2\theta_{13}=0.015$ seems to be
somewhat lower than the experimental lower bound.
However, as seen in Table 3, our prediction on $\delta_{CP}$
gives $\delta_{CP}=-103^\circ$, which decreases the lower 
bound $0.03$ of the T2K result to $0.02$.
Besides, the Double CHOOZ experiment \cite{DCHOOZ11} has reported that 
$\sin^2 2\theta_{13} = 0.085 \pm 0.029 \pm 0.042$ at $68 \%$ CL.
The lower value is $\sin^2 2\theta_{13}=0.014$.
Therefore, we consider that the predicted value
$\sin^2 2\theta_{13}=0.015$ is yet not ruled out, 
although the status is considerably severe.

\noindent
(ii) It also predicts a reasonable value of 
$R_\nu \equiv \Delta m^2_{solar}/\Delta m^2_{atm} \sim 0.03$ 
in contrast to the case of the O(3) model 
in which we could not predict $R_\nu$.  
(In the previous model, the value of $R_\nu$ needed to adjust 
the additional free parameter $m_{0\nu}^{-1}$ in Eq.(1.4).)

\noindent
(iii) The present model gives approximately degenerate neutrino masses
$m_{\nu 1} \sim m_{\nu 2}$.
The predicted value for the effective Majorana mass 
$\langle m_{ee}\rangle \simeq 0.0039$ eV
in the neutrinoless double beta decay  will be within our reach 
of the future experiments.

The big ansatz is the existence of the $X_2$
term in the charged lepton sector (1.12).
At present, there is no idea on this term.
Besides, it seems that the present lepton mass 
structure is ill matched with the charged lepton 
mass relation \cite{Koidemass} 
$$
\frac{m_e+m_\mu+m_\tau}{(\sqrt{m_e}+\sqrt{m_\mu}+\sqrt{m_\tau})^2}
=\frac{2}{3} .
\eqno(4.1)
$$
The purpose in the early stage of the yukawaon model was
to predict the charged lepton mass relation (4.1).
The bilinear form (1.2) for the charged lepton mass matrix was  
indispensable to predict \cite{Koide90MPL} the relation (4.1). 
If we adopt the present scenario, we must reconsider 
the origin of the charged lepton mass spectrum.
However, in this paper, we do not use the relation (4.1),
but only use the observed charged lepton mass values as 
input values. 
Therefore the bilinear form such as (1.2) 
is not necessarily required in this paper. 
Nevertheless, the formula (4.1) is still attractive.
On the other hand, it is also attractive that we can predict
12 observables (2+2+2 lepton and up-quark mass ratios,
and 4+2 PMNS mixing parameters) under 4 adjustable 
parameters $(a_e, a_u, \alpha_u, \zeta_u)$ if once we 
accept this ansatz (1.12). 
It is a future task how to understand the existence of 
$X_2$ term.

In conclusion, although the form $M_\nu$ is, at present, 
not one which is 
derived from a rigid theoretical ground, the form will 
offer a suggestive hint for a unification model of quark 
and lepton mass matrices.

\vspace{10mm}
{\Large\bf Acknowledgment}   

One of the authors (YK) is supported by JSPS 
(No.\ 21540266).

\newpage

\begin{center}
{\Large\bf \ Appendix A: $R$ charge assignments}
\end{center}

In the present model, as well as in the O(3) model, we construct a model 
without introducing a yukawaon
$Y_\nu$   by replacing $Y_\nu$ by $Y_e$. 
The simple way to guarantee that the yukawaon $Y_e$ couples not
only to the charged lepton sector but also to the Dirac 
neutrino sector is to introduce the following $R$ charge assignment, 
$$
R(\nu^c) = R(e^c) \equiv r_e, 
\eqno(A.1)
$$
$$
R(H_u) = R(H_d) = 1. 
\eqno(A.2)
$$  
The $R$ charge of $(\bar{E}_u E^u)$ is free parameter in the form (2.8). 
For simplicity, we take
$$
R(\bar{E}_u E^u) = R(\Theta_{8+1}) = 1. 
\eqno(A.3)
$$
Hereafter, we will denote $R(\bar{E}_u)$ and $R(E^u)$ as $\bar{r}_E$ and 
$1-\bar{r}_E$, respectively.
Each yukawaon is distinguished from other yukawaons by the $R$ charges. 
If we define a parameter $n$ as
$$
n \equiv 2[ R(\bar{Y}_R) - R(\bar{Y}_e) ] , 
\eqno(A.4)
$$
then, we can express the $R$ charges of the other fields from Eq.(2.2) 
as follows:
$$
R(\ell) = r_e + \frac{1}{2} (n - 2), 
\eqno(A.5)
$$
$$
R( \bar{Y}_e) = \frac{1}{2} (4 - n) - 2r_e , 
\eqno(A.6)
$$
$$
R(\bar{Y}_R) = 2 - 2 r_e, 
\eqno(A.7)
$$
$$
R(\bar{Y}_u) = n - 1 + \bar{r}_E, 
\eqno(A.8)
$$
$$
R(\bar{Y}_d) = \frac{1}{2} (n - 2) + \bar{r}_E, 
\eqno(A.9)
$$
$$
R(u^c) + R(q) = 2 - n - \bar{r}_E , 
\eqno(A.10)
$$
$$
R(d^c) + R(q) = 2 - \frac{1}{2} n - \bar{r}_E . 
\eqno(A.11)
$$
From Eqs.(2.6) and (2.7), we obtain
$$
R(\bar{Y}_u) = 2 R(\bar{\Phi}_u) + R(E^u), 
\eqno(A.12)
$$
$$
R(\bar{Y}_R) = R(\bar{\Phi}_u) + R(\bar{Y}_e) + R(E^u), 
\eqno(A.13)
$$
respectively. From Eqs.(A.12) and (A.13), we obtain a relation
$$
R(\bar{Y}_u) = n - R(E^u) = n - 1 + \bar{r}_E. 
\eqno(A.14)
$$
The relation (A.14) leads to 
$$
R(\bar{Y}_u \Theta^u) = (n - 1)R(\bar{E}_u E^u) + R(\bar{E}_u \Theta^u).  
\eqno(A.15)
$$
Only when the value $n$ is a positive integer, Eq.(A.15) means that
an additional term
$$
\frac{\lambda_{0u}}{\Lambda^{2n-1}} {\rm Tr}[(\bar{E}_u E^u)^{n-1} 
\bar{E} \Theta^u], 
\eqno(A.16)
$$
can appear in the expression (2.6). 
Note that if $n$ is not a positive integer, the factor 
$(\bar{E}_u E^u)^{n-1}$ does not have a physical meaning, 
because a term with $(E^u)^{-1}$ cannot appear in the 
superpotential terms. 
Therefore, the $n$ defined in Eq.(A.4) is allowed only for
$n=1,2,\cdots$. 

As we see in Eqs.(A.5) - (A.11), these $R$ charges are described 
by four parameters $r_e$, $R(q)$, $\bar{r}_E$ and $n$. 
Therefore, in order to fix these $R$ charge values, we have to 
assume four constraints for these $R$ charges. 
On the other hand, the fields $\bar{Y}_e$,  $\bar{Y}_R$, $\bar{Y}_u$, 
$\bar{Y}_d$, $\bar{\Phi}_u$, and $\bar{E}_u$ are gauge singlets, 
so that they must  be distinguished only by $R$ charges.
We can choose a suitable parameter set $(n, r_e, r_q, \bar{r}_E)$.
Here, let us demonstrate an example of $R$ charge assignments,
although it is not the purpose of the present paper to give such 
an explicit $R$ charge assignment.

For example, we put the following working hypothesis: 
$$
R(\bar{Y}_\nu) + R(\bar{Y}_e) = 0, \ \ \ 
R(\bar{Y}_u) + R(\bar{Y}_d) =  0, 
\eqno(A.17)
$$
$$
R(u^c) + R(d^c) = 0, \ \ \ R(\nu^c) + R(e^c) =  0. 
\eqno(A.18)
$$
The constraint (A.17) is an analogy that the Yukawa coupling
constants in the standard model do not have $R$ charges. 
The constraints (A.17) and (A.18) leads to the relation 
$R(\ell)=R(q)=1$.
Of course, since the yukawaon $\bar{Y}_\nu$ has been replaced by 
$\bar{Y}_e$ in the present model, the first constraint 
in Eq.(A.17) reads as $R(\bar{Y}_e)=0$, and since $R(\nu^c)=R(e^c)$ in
the model, the second constraint in Eq.(A.6) reads as $R(e^c)=0$.
Since $R(\bar{Y}_e)$ is given by Eq.(A.7), the requirement 
$R(\bar{Y}_e)=0$ together with $R(e^c)=0$ requires $n=4$. 
Thus, the constraints (A.17) and (A.18) fix the parameters
$(n, r_e, r_q, \bar{r}_E)$ as 
$$
n=4, \ \ R(e^c) =0 , \ \ R(q)=1, \ \ R(\bar{E}_u)=-2 .
\eqno(A.19)
$$
The explicit values of these $R$ values are listed in Table 1.
Since the $R$ charges of $\bar{\Phi}_0$, $X'$ and $X$ are still
free parameters, we take $R(\bar{\Phi}_0) =\frac{1}{2}$ for simplicity.
As we see in Table 1, the fields $\bar{Y}_e$, $\bar{Y}_R$, 
$\bar{Y}_u$, $\bar{Y}_d$ and $\bar{E}_u$ can safely have different $R$ 
charges from each other. 

Thus, the assumption can lead to plausible $R$ charge values (A.19), 
so that we consider that the assumption is reasonable.
Now we have an additional term, 
$$
   \frac{\lambda_{0u}}{\Lambda^{5}} {\rm Tr} [
  (\bar{E}_u E^u)^{3} \bar{E}_u \Theta^u ] ,
  \eqno(A.20)
$$ 
which should be included in $M_u$ given in Eq.(2.6). 

\newpage

\begin{center}
{\Large\bf \ Appendix B: Rotation from $S_3$ into $S_2$}
\end{center}

We define 
$$
S_3 =\frac{1}{3} \left(
\begin{array}{ccc}
1 & 1 & 1 \\
1 & 1 & 1 \\
1 & 1 & 1 
\end{array} \right) , \ \ \ \ 
S_2 =\frac{1}{2} \left(
\begin{array}{ccc}
1 & 1 & 0 \\
1 & 1 & 0 \\
0 & 0 & 0 
\end{array} \right) ,
\eqno(B.1)
$$
which are invariant under permutation symmetries S$_3$ and $S_2$,
respectively.
A rotation matrix $R$ which transforms the matrix $S_3$ into
$S_2$ has been discussed in Ref.\cite{KF02PRD}.
The rotation matrix $R$ is given by 
$$
R S_3 R^T = S_2 ,
\eqno(B.2)
$$
where $R$ is defined as follows:
$$
R =R_3(-\frac{\pi}{4}) \, T \, R_3(\theta)\, A ,
\eqno(B.3)
$$
$$
R_3(\theta) = \left(
\begin{array}{ccc}
\cos\theta & \sin\theta & 0 \\
-\sin\theta & \cos\theta & 0 \\
0 & 0 & 1
\end{array} \right),
\eqno(B.4)
$$
$$
T=\left(
\begin{array}{ccc}
0 & 0 & 1 \\
0 & 1 & 0 \\
1 & 0 & 0
\end{array} \right),
\eqno(B.5)
$$
$$
A=\left(
\begin{array}{ccc}
\frac{1}{\sqrt2} & -\frac{1}{\sqrt2}  & 0 \\
\frac{1}{\sqrt6}  & \frac{1}{\sqrt6} & -\frac{2}{\sqrt6} \\
\frac{1}{\sqrt3} & \frac{1}{\sqrt3} & \frac{1}{\sqrt3}
\end{array} \right) .
\eqno(B.6)
$$
The matrix $A$ is known as a matrix which diagonalizes
the matrix $S_3$ into
$$
A\, S_3 \, A^T = \left(
\begin{array}{ccc}
0 & 0 & 0 \\
0 & 0 & 0 \\
0 & 0 & 1
\end{array} \right) \equiv Z_3 ,
\eqno(B.7)
$$
and also
$$
TA\, S_3 \, (TA)^T = \left(
\begin{array}{ccc}
1 & 0 & 0 \\
0 & 0 & 0 \\
0 & 0 & 0
\end{array} \right) \equiv Z_1 .
\eqno(B.8)
$$

The explicit form of $R$ is given by
$$
R = \left(
\begin{array}{ccc}
\frac{1}{\sqrt6} - \frac{c}{2 \sqrt3} +\frac{s}{2}  & 
\frac{1}{\sqrt6} - \frac{c}{2 \sqrt3} -\frac{s}{2} & 
\frac{1}{\sqrt6} + \frac{c}{\sqrt3}  \\
\frac{1}{\sqrt6} + \frac{c}{2 \sqrt3} -\frac{s}{2}  & 
\frac{1}{\sqrt6} + \frac{c}{2 \sqrt3} +\frac{s}{2} & 
\frac{1}{\sqrt6} - \frac{c}{\sqrt3}  \\
\frac{c}{ \sqrt2} +\frac{s}{\sqrt6}  & 
- \frac{c}{\sqrt2} +\frac{s}{\sqrt6} & - \frac{2s}{\sqrt6}  
\end{array} \right) ,
\eqno(B.9)
$$
where $s=\sin\theta$ and $c=\cos\theta$.

In the expression (B.9) of $R$, when we define 
$$
\begin{array}{l}
z_1 = \frac{1}{\sqrt6} - \frac{c}{2 \sqrt3} +\frac{s}{2} , \\
z_2 = \frac{1}{\sqrt6} - \frac{c}{2 \sqrt3} -\frac{s}{2} , \\
z_3 = \frac{1}{\sqrt6} + \frac{c}{\sqrt3}  ,
\end{array}
\eqno(B.10)
$$
the rotation matrix $R$ is expressed as follows:
$$
R = \left(
\begin{array}{ccc}
z_1 & z_2 & z_3 \\
\sqrt{\frac{2}{3}} -z_1 &  \sqrt{\frac{2}{3}} -z_2 & 
\sqrt{\frac{2}{3}} -z_3 \\
\sqrt{\frac{2}{3}}(z_3 -z_2) & \sqrt{\frac{2}{3}}(z_1 -z_3) &
\sqrt{\frac{2}{3}}(z_2 -z_1) 
\end{array} \right) .
\eqno(B.11)
$$
Here, $z_i$ ($i=1,2,3$) satisfies 
$$
z_1^2 + z_2^2 + z_3^2 =1 ,
\eqno(B.12)
$$
and we can choose $z_i$ such as 
$$
z_1 + z_2 +z_3 = \sqrt{\frac{3}{2}} .
\eqno(B.13)
$$
Suggested from the charged lepton mass relation (4.1), 
if we choose $z_i$ as
$$
z_i = \frac{\sqrt{m_{ei}}}{\sqrt{m_{e1}+m_{e2}+m_{e3}}} ,
\eqno(B.14)
$$
where $(m_{e1}, m_{e2}, m_{e3})=(m_e, m_\mu, m_\tau)$, 
then, the matrix $R$ satisfies 
$$
R \left(
\begin{array}{c}
z_1 \\
z_2 \\
z_3
\end{array} \right) = \left(
\begin{array}{c}
1 \\
0 \\
0
\end{array} \right) .
\eqno(B.15)
$$
Since
$$
Z S_3 Z = \frac{1}{3} \left(
\begin{array}{ccc}
z_1^2 & z_1 z_2 & z_1 z_3 \\
z_1 z_2 & z_2^2 & z_2 z_3 \\
z_1 z_3 & z_2 z_3 & z_3^2 
\end{array} \right) ,
\eqno(B.16)
$$
where 
$$
Z = \left( 
\begin{array}{ccc}
z_1 & 0 & 0 \\
0 & z_2 & 0 \\
0 & 0 & z_3
\end{array} \right) ,
\eqno(B.17)
$$
the following relation holds:
$$
R Z S_3 Z R^T = \frac{1}{3} Z_1 ,
\eqno(B.18)
$$
where $Z_1$ is defined by Eq.(B.8).
However, note that, from Eqs.(B.8) and (B.18), we cannot 
conclude $R Z = (1/\sqrt3) T A$.

Thus, it seems the rotation matrix $R$ from $S_3$ into $S_2$ 
is deeply related to the charged lepton mass relation (4.1),
but it is not clear why the form $S_2$ appears in the charge lepton
sector. 
This is still an open question at present.

\newpage


\end{document}